\documentclass[prd,superscriptaddress,twocolumn,nofootinbib,showpacs,preprintnumbers,floatfix]{revtex4}

\usepackage{mathrsfs}
\usepackage{amsfonts,amssymb,amsmath}
\usepackage{graphicx,epsfig}
\usepackage{multirow}
\usepackage{verbatim}

\def\fsu5{$\cal{F}$-$SU(5)$}
\def\bfsu5{$\boldsymbol{\mathcal{F}}$-$\boldsymbol{SU(5)}$}

\def\m1half{$M_{1/2}$}
\def\m3half{$M_{3/2}$}
\def\m32{$M_{32}$}

\def\fb{${\rm fb}^{-1}$~}

\def\mt2{$M_{T2}$}
\def\x2{$\chi^2$}
\def\2b{$M_{T2}b$}
\def\sb{$S/\sqrt{B+1}$~}
\def\bs0{$B_S^0 \rightarrow \mu^+ \mu^-$}

\begin{document}

\title{Testing No-Scale \fsu5:\\ A 125 GeV Higgs Boson and SUSY at the $\sqrt{s}$ = 8 TeV LHC}

\author{Tianjun Li}

\affiliation{State Key Laboratory of Theoretical Physics and Kavli Institute for Theoretical Physics China (KITPC),
Institute of Theoretical Physics, Chinese Academy of Sciences, Beijing 100190, P. R. China}

\affiliation{George P. and Cynthia W. Mitchell Institute for Fundamental Physics and Astronomy, Texas A$\&$M University, College Station, TX 77843, USA}

\author{James A. Maxin}

\affiliation{George P. and Cynthia W. Mitchell Institute for Fundamental Physics and Astronomy, Texas A$\&$M University, College Station, TX 77843, USA}

\author{Dimitri V. Nanopoulos}

\affiliation{George P. and Cynthia W. Mitchell Institute for Fundamental Physics and Astronomy, Texas A$\&$M University, College Station, TX 77843, USA}

\affiliation{Astroparticle Physics Group, Houston Advanced Research Center (HARC), Mitchell Campus, Woodlands, TX 77381, USA}

\affiliation{Academy of Athens, Division of Natural Sciences, 28 Panepistimiou Avenue, Athens 10679, Greece}

\author{Joel W. Walker}

\affiliation{Department of Physics, Sam Houston State University, Huntsville, TX 77341, USA}


\begin{abstract}
We celebrate the recent Higgs discovery announcement with our experimental colleagues at the LHC and look forward
to the implications that this success will bring to bear upon the continuing search for supersymmetry (SUSY).
The model framework named No-Scale \fsu5 possesses the rather unique capacity to provide 
a light CP-even Higgs boson mass in the favored 124--126 GeV window while simultaneously retaining
a testably light SUSY spectrum that is consistent with emerging low-statistics excesses
beyond the Standard Model expectation in the ATLAS and CMS multijet data.
In this letter we review the distinctive \fsu5 mechanism that forges the physical 125 GeV Higgs boson and  
make a specific assessment of the ATLAS multijet SUSY search observables that may be expected for a 15 \fb delivery
of 8 TeV data in this model context.  Based on our Monte Carlo study, we anticipate that the enticing hints of a SUSY signal
observed in the 7 TeV data could be amplified in the 8 TeV results.  Moreover, if the existing signal
is indeed legitimate, we project that the rendered gains in significance will be sufficient to conclusively rule out an
alternative attribution to statistical fluctuation at that juncture.
\end{abstract}



\preprint{ACT-10-12, MIFPA-12-27}

\maketitle


We take the occasion of the historic July 4, 2012 joint announcement~\cite{atlas:report,cms:report} by the Large Hadron
Collider (LHC) ATLAS and CMS collaborations regarding the status of the Higgs boson search to offer a most
heartfelt congratulations to our experimental colleagues, including the full host of physicists and technicians
tasked with maintaining an efficient and stable beam operation environment, detecting and recording the intricate
visible signature of each unique event, distributing and processing innumerable Terabytes of raw information, and reconstructing the
delicate natural order that collectively underlies the violence and chaos of each isolated collision.
As this grand effort now reaches the five standard deviation statistical threshold that by
conventional assent represents ``discovery'', we stand with the rest of the world's scientific community
to raise a glass.

Together with this moment of celebration arrives also an opportunity for reflection on the fate to come
for the second great LHC search target of supersymmetry (SUSY).  Such contemplation must take to account
the existing state of decimation exacted upon the leading minimal supergravity (mSUGRA) and constrained
minimal supersymmetric standard model (CMSSM) constructions by the 2011 5~\fb data integration at $\sqrt{s} = 7$~TeV,
the projected consequences for the collected but unstudied 6~\fb of data at $\sqrt{s} = 8$~TeV, and perhaps
most importantly, the new constraints imposed by the emerging reality of a light Higgs boson at a mass around 125~GeV.
It is for these causes that we take pen to hand for a short letter regarding our best anticipations
for the next great LHC year, refocused by the successes of the year passed.

It is only through the specificity afforded by a model that one may hope to correlate the
Higgs and SUSY searches, or to make projections for the expected yield of various event selection strategies
to be applied against the present accumulation of data.  We know only one particular class of model
possessing the capacity to simultaneously satisfy constraints related to the dark matter relic density,
rare physical processes, precision electroweak measurements, LHC squark and gluino searches, and now also,
the light Higgs boson mass.  Satisfying the dynamically established boundary conditions of No-Scale supergravity~\cite{Cremmer:1983bf,Ellis:1983sf, Ellis:1983ei, Ellis:1984bm, Lahanas:1986uc} and
featuring the field content of the Flipped $SU$(5) grand unified theory~\cite{Barr:1981qv,Derendinger:1983aj,Antoniadis:1987dx} (GUT) with additional TeV-scale vector-like
supersymmetric multiplets~\cite{Jiang:2006hf,Jiang:2009zza,Jiang:2009za,Li:2010dp,Li:2010rz} (flippons) (for previous
study, see~\cite{Moroi:1991mg,Babu:2004xg}), these constructions have been dubbed No-Scale \fsu5.  This family
of models has not only survived the onslaught of negative LHC data, but has unambiguously demonstrated
the ability to handily produce a 125~GeV Higgs boson while efficiently explaining emergent low-statistics
positive excesses above Standard Model (SM) expectations in the elusive SUSY hunt.  Crucially, this congruence with experiment
has been achieved while reducing the overall cardinality of parameterization available to the CMSSM, opting
instead for the enhanced parsimony, predictive stricture and depth of theoretical motivation inherent to
the No-Scale framework.  We shall review the unique mechanisms employed by \fsu5 toward the realization of these physical
observables, and project forward how SUSY might subsequently join the Higgs boson as an experimentally detected phenomenon.

No-Scale \fsu5 (See
Refs.~\cite{Li:2010ws,Maxin:2011hy,Li:2011hr,Li:2011xu,Li:2011rp,Li:2011fu,Li:2011xg,Li:2011av,Li:2011ab,Li:2012hm,Li:2012tr,Li:2012ix,Li:2012yd}
and all references therein) is constructed upon the foundation of the Flipped $SU$(5)~\cite{Barr:1981qv,Derendinger:1983aj,Antoniadis:1987dx} GUT, two pairs of
hypothetical TeV-scale flippons of mass $M_V$ derived from local F-Theory model building~\cite{Jiang:2006hf,Jiang:2009zza,Jiang:2009za,Li:2010dp,Li:2010rz}, and the
dynamically established boundary conditions of No-Scale supergravity~\cite{Cremmer:1983bf,Ellis:1983sf, Ellis:1983ei, Ellis:1984bm, Lahanas:1986uc}.  In the simplest No-Scale scenario,
$M_0$=A=$B_{\mu}$=0 at the unification boundary, while the complete collection of low energy SUSY breaking soft-terms evolve down
with a single non-zero parameter $M_{1/2}$.  Consequently, the particle spectrum will be proportional to $M_{1/2}$ at leading order,
rendering the bulk ``internal'' physical properties invariant under an overall rescaling.  The matching condition between the low-energy value of
$B_\mu$ that is demanded by EWSB and the high-energy $B_\mu = 0$ boundary is notoriously difficult to reconcile under the
renormalization group equation (RGE) running.  The present solution relies on modifications to the $\beta$-function coefficients that are generated by the flippon
loops.  Naturalness in view of the gauge hierarchy and $\mu$ problems suggests that the flippon mass $M_{\rm V}$ should be of the TeV order.
Avoiding a Landau pole for the strong coupling constant restricts the set of vector-like multiplets which may be
given a mass in this range to only two constructions with flipped charge assignments, which have been explicitly realized
in the $F$-theory model building context~\cite{Jiang:2006hf,Jiang:2009zza, Jiang:2009za}.  In either case, the (formerly negative) one-loop $\beta$-function
coefficient of the strong coupling $\alpha_3$ becomes precisely zero, flattening the RGE running, and generating a wide
gap between the large $\alpha_{32} \simeq \alpha_3(M_{\rm Z}) \simeq 0.11$ and the much smaller $\alpha_{\rm X}$ at the scale $M_{32}$ of the intermediate
flipped $SU(5)$ unification of the $SU(3)_{\rm C} \times SU(2)_{\rm L}$ subgroup.  This facilitates a very significant secondary running phase
up to the final $SU(5) \times U(1)_{\rm X}$ unification scale, which may be elevated by 2-3 orders of magnitude
into adjacency with the Planck mass, where the $B_\mu = 0$ boundary condition fits like hand to glove~\cite{Ellis:2001kg,Ellis:2010jb,Li:2010ws}.

\begin{figure*}[htp]
        \centering
        \includegraphics[width=1.00\textwidth]{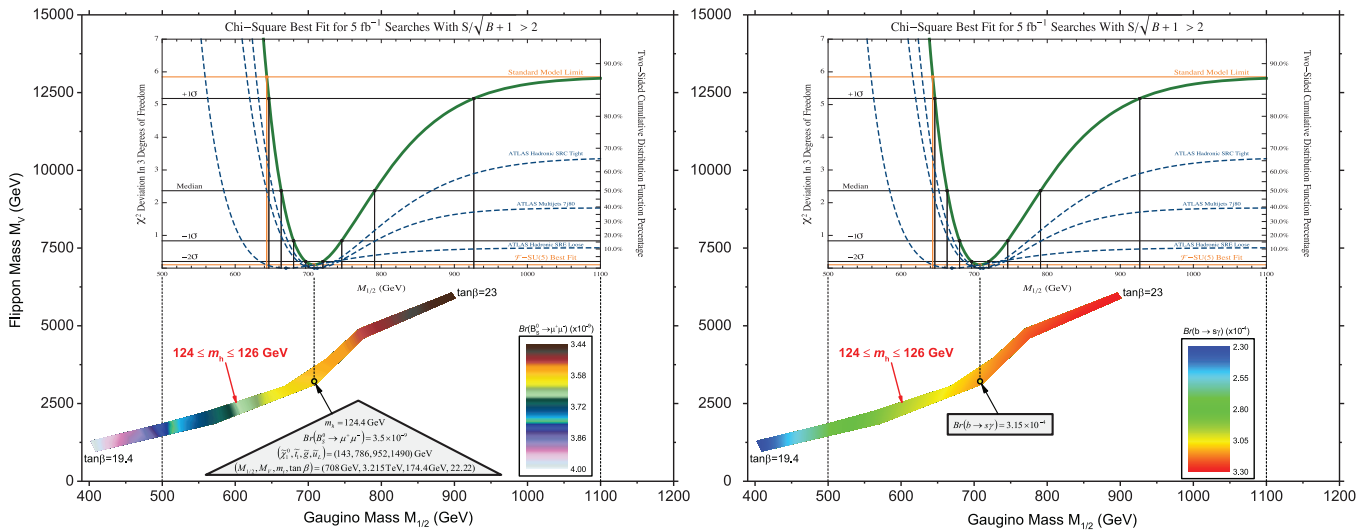}
        \caption{We depict the experimentally viable parameter space of No-Scale \fsu5 as a function of the gaugino mass $M_{1/2}$
and flippon mass $M_V$. The surviving model space after application of the bare-minimal constraints of
Ref.~\cite{Li:2011xu}
and Higgs boson mass calculations of Ref.~\cite{Li:2011ab} is illustrated by the narrow strip with the
smoothly contoured
color gradient. The gradient represents the total branching ratio (SM+SUSY) of the B-decay process \bs0
(left), and the total branching ratio (SM+SUSY) of $b \to s \gamma$ (right). The inset diagrams (with
linked horizontal scale) are the multi-axis cumulative \x2
fitting of Ref.~\cite{Li:2012tr}, depicting the best SUSY mass fit and Standard Model limit of only those
ATLAS and CMS
SUSY searches exhibiting a signal significance of \sb $>$ 2. The best fit benchmark of
Ref.~\cite{Li:2012tr} is highlighted at $M_{1/2}$ = 708 GeV, with $m_h$ = 124.4 GeV.}
        \label{fig:sliver}
\end{figure*}

The \fsu5 model space is bounded primarily by a set of ``bare-minimal'' experimental constraints distinguished 
by a great longevity of relevance, as defined in Ref.~\cite{Li:2011xu}. These include the top quark mass
$172.2~{\rm GeV} \leq m_{\rm t} \leq 174.4~{\rm GeV}$,
7-year WMAP cold dark matter relic density $0.1088 \leq \Omega_{\rm CDM}h^2 \leq
0.1158$~\cite{Komatsu:2010fb}, and precision LEP constraints on the SUSY mass content. We further
append to this classification an adherence to the defining high-scale boundary conditions of the model.
In light of recent developments, the favored parameter space may be further
circumscribed by the demands of a 124--126 GeV Higgs boson mass.  The surviving region is comprised
of a narrow strip of space confined to 400 $\le M_{1/2} \le$ 900 GeV, 19.4 $\le$ tan$\beta$ $\le$ 23, and 950
$\le M_V \le$ 6000 GeV, as illustrated in Figure (\ref{fig:sliver}). The border at the minimum $M_{1/2}$ =
400 GeV is required by the LEP constraints, while the maximum boundary at $M_{1/2}$ = 900 GeV prevents a
charged stau LSP at around tan$\beta \cong$ 23.  In the bulk of the model space the lightness of the stau,
which is itself a potential future target for direct collider probes by the forthcoming $\sqrt{s} = 14$~TeV LHC,
is leveraged to facilitate an appropriate dark matter relic density via stau-neutralino coannihilation.
The SUSY particle masses and relic densities are
calculated with {\tt MicrOMEGAs 2.1}~\cite{Belanger:2008sj}, via application of a proprietary
modification of the {\tt SuSpect 2.34}~\cite{Djouadi:2002ze} codebase to evolve the flippon-enhanced RGEs.

The convergence of the predicted \fsu5 Higgs mass with the collider measured value
is achieved through contributions to the lightest CP-even Higgs boson mass from the flippons, calculated from the
RGE improved one-loop effective Higgs potential approach~\cite{Babu:2008ge,Martin:2009bg}. The
mechanism for the serendipitous mass shift is a pair of Yukawa interaction terms between the Higgs and
vector-like flippons in the superpotential, resulting in a 3--4 GeV upward shift in the Higgs mass to the
experimentally measured range~\cite{Li:2011ab}. Using the relevant shift in the Higgs mass-square
as approximated in Refs.~\cite{Huo:2011zt,Li:2011ab}, which implements a leading dependence of the
flippon mass $M_V$, larger shifts correspond to lighter vector-like flippons. This flippon induced
mechanism operates in synthesis with the top quark mass, whose elevation similarly raises the
non-flippon contributed Higgs mass. The cumulative result is the very narrow strip of model space in
Figure (\ref{fig:sliver}), with the lower strip boundary truncated by the upper top quark mass
extremity, and the upper strip boundary situated at the minimum Higgs mass of 124 GeV, conveniently
establishing a stable, thin band of experimentally viable points with which to explore new physics.

A pair of key venues for the appearance of new physics are the rare decay processes \bs0 and $b \to s \gamma$. In
Ref.~\cite{Li:2012yd}, we analyzed the recently improved LHCb constraints on the flavor changing
neutral current B-decay process \bs0. It was discovered that the SUSY contribution throughout the narrow
strip of model space is much smaller than the computed SM prediction of $Br$(\bs0)~$ = (3.2 \pm 0.2) \times
10^{-9}$~\cite{Buras:2010mh,Buras:2010wr}, in fine consistency with the LHCb result of
$Br$(\bs0)~$< 4.5(3.8) \times 10^{-9}$ at the 95\% (90\%) confidence level~\cite{Aaij:2012ac}. The
left-hand plot space of Figure (\ref{fig:sliver}) illustrates smoothly graded contours of color
depicting the calculation of $Br$(\bs0) within the model space, using {\tt MicrOMEGAs 2.1} and our
proprietary modification of {\tt SuSpect 2.34}. For the benchmark highlighted, which is favored by the low-statistics
SUSY event over-production already observed in the 5 \fb 7 TeV data to be discussed shortly, the
$Br(B^0_s \to \mu^+ \mu^-) = 3.5 \times 10^{-9}$ is very close to the SM prediction, thus suggesting that indeed the
\fsu5 SUSY contribution is quite small, consistent with indications from experiment that any SUSY
contribution must be a great deal less than the SM value. We have suppressed here the flippon contributions
by appealing to some combined effect of the natural heaviness of the multiplets, and
an assumption that the mixings between the flippons and the SM fermions are relatively small.

The other rare decay process to which we extend a detailed analysis is $b \to s \gamma$. The right-hand plot
space of Figure (\ref{fig:sliver}) exhibits smooth color gradients of $Br(b \to s \gamma)$, computed
with {\tt MicrOMEGAs 2.4}~\cite{Belanger:2010gh}, where the SM and Higgs contributions at NLO are
included, in addition to the leading order SUSY contributions. The emphasized benchmark computes to
$Br(b \to s \gamma) = 3.15 \times 10^{-4}$, which is noteworthy for its quite close proximity to the SM
estimate. The NNLO SM contribution was computed to be $Br(b \to s \gamma) = (3.15 \pm 0.23) \times
10^{-4}$~\cite{Misiak:2006zs,Misiak:2006ab}, though it was shown that this SM value could be
reproduced by the NLO calculation by choice of the c-quark mass
scale~\cite{Misiak:2006ab,Gambino:2008fj}. As such, the {\tt MicrOMEGAs 2.4} codebase, employed
here, computes the NLO SM contribution at $Br(b \to s \gamma) = 3.27 \times 10^{-4}$, corresponding to the
result of Ref.~\cite{Gambino:2008fj}. Therefore, as shown in Figure (\ref{fig:sliver}), our \fsu5
benchmark result of $Br(b \to s \gamma) = 3.15 \times 10^{-4}$ includes only a very small negative SUSY
contribution to the SM estimate, though well within theoretical uncertainty and experimental
resolution.

The intriguing aspect of the \fsu5 contribution to the rare decay processes is the nearly negligible SUSY
contributions to both \bs0 and $b \to s \gamma$. The benchmark model denoted in Figure (\ref{fig:sliver})
carries with it virtually no SUSY contribution to these two processes. Hence, in an era when evidence supporting
any sizable contribution to these rare decay processes by new physics is rapidly fading away
and consequently disfavoring a majority of SUSY constructions, the experimental standing of No-Scale \fsu5
is instead characteristically enhanced.  With the \fsu5 SUSY contributions positioned interior to any
foreseeable uncertainty bounds around the SM value, it is difficult to envision a scenario under
which \fsu5 could suffer exclusion by these rare decay processes, absent an unexpected reversion of the
experiments toward the side of a large SUSY contribution.  Topically parallel to this discussion
is the possibility of post-SM contributions to the magnetic moment $(g-2)_{\mu}$ of the muon, although we 
extend somewhat less credulity to the related limits due to large lingering uncertainties in the QCD calculations,
an inconclusive experimental deviation from zero, and large discrepancies between the $e^+e^-$
and $\tau^+ \tau^-$ based results.  Nevertheless, the \fsu5 SUSY contribution $\Delta a_{\mu}$ follows
the lead of \bs0 and $b \to s \gamma$ in this regard, where $\Delta a_{\mu} \cong 7.5 \times 10^{-10}$ for
the benchmark as computed with {\tt MicrOMEGAs 2.4} is again only a very small deviation from the SM estimate,
and within any anticipated future experimental uncertainty. Therefore, we observe a comforting
consistency amongst these three rare decay processes in \fsu5 through each of their practically
negligible SUSY contributions.

The same flippon induced perturbation to the RGE unification structure of \fsu5 that was responsible
for facilitating a consistent application of the No-Scale boundary conditions near the Planck mass
also produces a key phenomenological signature.  The flat RGE evolution of the $SU(3)_C$ gaugino mass $M_3$, which 
mirrors the flatness of the $\beta$-coefficient $b_3 = 0$, suppresses the standard logarithmic mass enhancement at low-energy
and yields a SUSY spectrum $M(\widetilde{t}_1) < M(\widetilde{g}) < M(\widetilde{q})$
where the light stop $\widetilde{t}_1$ and gluino $\widetilde{g}$ are both less massive than all other squarks.
This highly unusual hierarchy produces a distinct event topology initiated by the pair-production
of heavy first or second generation squarks $\widetilde{q}$ and/or gluinos in the hard scattering
process, with the heavy squark likely to yield a quark-gluino pair $\widetilde{q} \rightarrow q \widetilde{g}$.
The gluino then has only two main channels available in the cascade decay,
$\widetilde{g} \rightarrow \widetilde{t}_1 \overline{t}$ or $\widetilde{g} \rightarrow q \overline{q} \widetilde{\chi}_1^0$,
with $\widetilde{t}_1 \rightarrow t \widetilde{\chi}_1^0$ or $\widetilde{t}_1 \rightarrow b \widetilde{\chi}_1^{\pm}$.
As $M_{1/2}$ increases, the stop-top channel becomes
dominant, ultimately reaching 100\% for $M_{1/2} \ge 729$ GeV. For $M_{1/2} < 729$ GeV, both avenues have
sufficient branching fractions to produce observable events at the LHC. Each gluino produces 2--6
hadronic jets, with the maximum of six jets realized in the gluino-mediated stop decay, so that a single gluino-gluino
pair-production event can net 4--12 jets.  After further fragmentation processes, the final event is
characterized by a definitive SUSY signal of high-multiplicity jets.

The most robust test of any supersymmetric model is the prediction of a unique signature plainly accounting
for observed anomalies in collider data.  The exceptional mass ordering in No-Scale \fsu5 provides a
distinctive marker at the LHC, since multijet events are expected to dominate a probed \fsu5 framework. We
first suggested in March 2011~\cite{Maxin:2011hy,Li:2011hr} that SUSY in an \fsu5 universe would become
manifest at the colliders in high-multiplicity jet events, extending this initial study in
Refs.~\cite{Li:2011rp,Li:2011fu,Li:2011xg,Li:2011av,Li:2011ab,Li:2012hm,Li:2012tr,Li:2012ix}.
The first ample accumulation of multijet data was released by the collaborations later in 2011 in
Refs.~\cite{PAS-SUS-11-003,Aad:2011ib,Aad:2011qa}, based upon 1 \fb of luminosity. Though the
number of events remaining after the collaboration data cuts was less than ten, there did
appear small but curious excesses beyond the SM estimates in these searches targeting multijet events.
The most prominent examples came from ATLAS, where the 7j80 ($\ge$ 7 jets and jet $p_T >$ 80 GeV) search of
Ref.~\cite{Aad:2011qa} and High Mass ($\ge$ 4 jets and jet $p_T >$ 80 GeV) search of Ref.~\cite{Aad:2011ib}
displayed interesting event production over the data-driven background estimates. Employing the signal significance
metric \sb, we computed a value of 1.1 for 7j80 and 1.3 for the High Mass search.  Despite the weak signal,
reasonably attributable to statistical fluctuations, No-Scale \fsu5 provided a neat and efficient explanation
for the minor over-productions in these two searches.  Despite the long odds at that time, those clean
fits prompted us to extrapolate from the ATLAS published statistics of Refs.~\cite{Aad:2011ib,Aad:2011qa}
to predict signal strengths of \sb = 1.9 for 7j80 and \sb = 3.0 for the High Mass~\cite{Li:2012hm} search
in the forthcoming 5 \fb data set at 7 TeV, assuming a legitimate physics origin for the intriguing over-production.

We provided a detailed analysis of the ATLAS and CMS 5 \fb observations at the 7 TeV LHC in
Ref.~\cite{Li:2012tr}, focused on those search strategies where the signal significance was strongest
and the largest number of events had accumulated, imposing \sb $>$ 2.0 as a minimal boundary. 
Strikingly, the 7j80~\cite{ATLAS-CONF-2012-037} search and the composite successors to the High Mass search were the only 5
\fb strategies to surmount this significance hurdle.
To elaborate, ATLAS essentially segregated the former High Mass $\ge$4 jet SUSY search of Ref.~\cite{Aad:2011ib} into three separate
searches of 4 jets, 5 jets, and 6 jets for the latter study, intended to isolate the $\widetilde{g}\widetilde{g}$,
$\widetilde{q}\widetilde{g}$, and $\widetilde{q}\widetilde{q}$ 0-lepton channels via $\widetilde{q}
\rightarrow q\widetilde{g}$ and $\widetilde{g} \rightarrow q
\overline{q} \widetilde{\chi}_1^0$~\cite{ATLAS-CONF-2012-033}. In addition to the ATLAS
7j80~\cite{ATLAS-CONF-2012-037}, these ATLAS 4-jet and 6-jet searches of
Ref.~\cite{ATLAS-CONF-2012-033}, referred to as SRC Tight and SRE Loose, respectively, were the only
other 5 \fb searches to achieve \sb $>$ 2.0 in all the ATLAS and CMS 5 \fb studies analyzed at that time.
Granting that the 1 \fb data sample is a subset of the 5 \fb data,
the signal strength nevertheless expanded in the precise proportionality expected.
The final 5 \fb 7 TeV ATLAS observations computed signal significances of \sb =
2.1 for 7j80~\cite{ATLAS-CONF-2012-037}, \sb = 3.2 for SRC Tight (4j)~\cite{ATLAS-CONF-2012-033},
and \sb = 2.6 for SRE Loose (6j)~\cite{ATLAS-CONF-2012-033}, in line with our predictions and very
consistent with the signal growth expected to be observed in an \fsu5 universe.

This enlarged signal strength simultaneously presented a golden opportunity to derive a best fit SUSY mass to the 5 \fb data
through a \x2 fitting procedure.  We demonstrated~\cite{Li:2012tr} clear internal consistency in the \fsu5 mass scale favored
by the various search windows, in addition to the described correlation across time in the signal growth.
This analysis favored sparticle masses of $m_{\widetilde{\chi}_1^0}$ = 143 GeV,
$m_{\widetilde{t}_1}$ = 786 GeV, $m_{\widetilde{g}}$ = 952 GeV, and $m_{\widetilde{u}_L}$ = 1490 GeV,
complementing a Higgs mass of $m_{h}$ = 124.4 GeV at the $M_{1/2}$ = 708 GeV well of the 5 \fb multi-axis
cumulative \x2 curve, combining the 7j80, SRC Tight, and SRE Loose search channels.
To exemplify this best fit at the \x2 minimum, we choose the $M_{1/2}$ = 708 GeV point as our standing favored
benchmark, exhibited and pinpointed in Figure (\ref{fig:sliver}).  The superimposed cumulative \x2
curve of Ref.~\cite{Li:2012tr} visibly showcases how the ATLAS 7j80, SRC Tight, and SRE Loose over-productive
search strategies illuminate the \fsu5 model space as naturally conforming to the collider observations.
By lowering the minimum threshold for signal significance to \sb $>$ 1.0, the CMS 5 \fb MT2 search strategy~\cite{SUS-12-002}
was included into our 5 \fb multi-axis \x2 fitting in Ref.~\cite{Li:2012ix} along with an additional ATLAS search, namely the 8j55
case from Ref.~\cite{ATLAS-CONF-2012-037}.  It was demonstrated in this manner that further non-trivial correlations exist between
the mass scale favored by independently productive ATLAS and CMS SUSY searches, bolstering the case against attribution 
of the excesses to random statistical fluctuations.

The data observations for the ATLAS multijet searches discussed here have shown a very
natural progression from 1 \fb to 5 ${\rm fb^{-1}}$. In fact, the \sb $\sim$ 3 signal significance of the
combined ATLAS 5 \fb multijet searches, which we can consider to be about 3$\sigma$, is near the same signal
level as the Higgs boson after 5 \fb at 7 TeV. With the Higgs boson now at the discovery threshold of 5$\sigma$
in the first 8 TeV data tranche, it would only be fitting if the ATLAS multijet SUSY searches
continued to track the Higgs signal strength. Looking forward and preparing for potentially more significant SUSY production as we shift to forthcoming larger LHC beam collision energies and hence greater numbers of statistics, we transition here to a more appropriate metric for measuring signal strength in the presence of larger excess event production beyond expectations, $2 \times (\sqrt{S+B} - \sqrt{B})$. We employ the background statistics derived by the ATLAS Collaboration for 5 \fb at 7 TeV from Ref.~\cite{ATLAS-CONF-2012-037}, though to determine an estimate of the SM background for 8 TeV, we scale up these ATLAS statistics using the same factor observed in our Monte Carlo for \fsu5 simulations. This estimator, while serving our limited scope here satisfactorily, can only be as reliable as the expectation of statistical, dynamic and procedural stability across the transition in energy, luminosity and model. We further assume here a static data cutting strategy between the ATLAS 7 and 8 TeV multijet searches. We indeed project that there should be a visible multijet signal strength sufficient for SUSY discovery in the isolated 15 \fb 8 TeV data, expected to be recorded in 2012 and processed in 2013, if the existing signal in the 5 \fb 7 TeV data is legitimately and wholly attributable to new physics. More precisely, assuming no important modifications to the background calibration procedures by ATLAS, we can project the 7j80 SUSY search tactics of Ref.~\cite{ATLAS-CONF-2012-037} to yield a signal significance of $2 \times (\sqrt{S+B} - \sqrt{B}) \sim 6$ for 15 \fb at 8 TeV, and $2 \times (\sqrt{S+B} - \sqrt{B}) \sim$ 7--8 for the SRC-Tight and SRE-Loose search strategies of Ref.~\cite{ATLAS-CONF-2012-033}. Although potentially quite susceptible to large statistical fluctuation, these rather strong signal projections nonetheless indicate that a probing of the \fsu5 framework at the LHC could indeed yield further tantalizing, and possibly convincing, evidence that nature herself is fundamentally supersymmetric. The summation of the 5 \fb of 7 TeV data to the 8 TeV data only improves the signal significance modestly. Moreover, the presence of excess events in the 15 \fb ATLAS multijet searches at 8 TeV will resoundingly indicate that random background anomalies are not the source of the 7 TeV multijet over-production. We find the predictable evolution of our SUSY exploration from the initial 1 \fb at 7 TeV to the 5 \fb at 7 TeV to warrant such positive speculation as we move forward to the 15 \fb at 8 TeV.

After taking pause to revel in the landmark discovery of the Higgs boson, we
arrive now at potentially the most consequential inflection point in the dual effort to reveal the 
definitive imprint of supersymmetry upon natural processes.  The most welcome news of a light Higgs around 125~GeV
serves ironically as yet another nail into the coffin of the most conventional mSUGRA/CMSSM constructions,
which become irreconcilably torn between two masters in the effort to pair this elevated scale with a
testably light SUSY spectrum.  The explanation of any persistent production excesses in multijet events will 
require a turn toward alternative formulations such as No-Scale \fsu5 for a proper accounting.  In such a
context, the empirically confirmed addition of the Higgs into the fold of known particles serves only to
further narrow the already precise model predictions.  Success builds upon success, and the next stride
into the 8~TeV SUSY search collision data could cement 2012 as a truly historic year for particle physics.


\begin{acknowledgments}
This research was supported in part
by the DOE grant DE-FG03-95-Er-40917 (TL and DVN),
by the Natural Science Foundation of China
under grant numbers 10821504, 11075194, and 11135003 (TL),
by the Mitchell-Heep Chair in High Energy Physics (JAM),
and by the Sam Houston State University
2011 Enhancement Research Grant program (JWW).
We also thank Sam Houston State University
for providing high performance computing resources.
\end{acknowledgments}


\bibliography{bibliography}

\end{document}